\documentstyle[12pt]{article}

\begin{document}
\setlength{\baselineskip}{24pt}
\begin{titlepage}
\begin{center}
{\bf  Relativistically  Covariant   Symmetry in QED}

\bigskip

{\centerline {Zhong Tang $^{*}$  and  David Finkelstein $^{**}$}}

\medskip

{\centerline {School of Physics, Georgia Institute of Technology}}
{\centerline {Atlanta, GA 30332-0430, USA}}
\end{center}

\vspace{.6cm}
{\centerline {abstract}}
\noindent
We construct a relativistically  covariant  symmetry
of QED. Previous  local and nonlocal symmetries are
special  cases. This generalized
symmetry  need not be nilpotent,
but  nilpotency   can be
arranged with an auxiliary field and a certain condition.
The Noether charge  generating the symmetry transformation
is obtained, and it imposes a constraint on the physical states.


\bigskip

E-mail addresses:* gt8822b@prism.gatech.edu,

**david.finkelstein@physics.gatech.edu

\end{titlepage}

\newpage

Quantum gauge theory  is founded on phase symmetry,
but gauge degrees of freedom  bring in  extra independent variables.
One introduces gauge conditions to suppress  these
 variables, but destroys the gauge symmetry thereby.
In path integral form, with the introduction of ghosts,
gauge invariance  is recovered through the passage to
the BRST cohomology \cite{s2}. The BRST  theory  raises the ghosts to
a prominent role for it regards all  fields, including
ghosts, as elements of a single geometrical object,
the cohomology.

 Since locality has been argued
to be the main cause of infinities in the usual quantum
field theory, people have been turning to  nonlocal quantum
field theory \cite{s3,s4}.  Nonlocal gauge symmetry
plays an important role in nonlocal quantum field theories.

Lavelle and McMullan's recent work \cite{s5} ingeniously reveal that
 local QED exhibits a nonlocal symmetry,  here called the LM symmetry,
  which is nilpotent
 but not Lorentz  covariant. They dealt  with the following Lagrangian
 with a gauge fixing term and ghosts $C(x), \bar{C}(x)$ \cite{s6}:
\begin{equation}
L=-\frac{1}{4}F_{\mu\nu}F^{\mu\nu}-{1\over{2\xi}}
(\partial_{\mu}A^{\mu})^{2}
+\bar{\psi}(i\gamma_{\mu}D^{\mu}-m)\psi+i\bar{C}\partial_{\nu}
\partial^{\nu}C,
\end{equation}
where $ D_{\mu}=\partial_{\mu}-ig_{0}A_{\mu} $.
The  nonlocal  LM transformation is
\newpage
\begin{equation}
\begin{array}{ll}
\delta A_{0}=i\bar{C},&  \delta A_{i}=i\frac{\partial_{i}\partial_{0}}
{\nabla^{2}}\bar{C}, \\[.15in]
\delta C=-A_{0}+\frac{\partial_{i}\partial_{0}}{\nabla^{2}}A_{i}+
\frac{g_{0}}{\nabla^{2}}\bar{\psi}\gamma_{0}\psi,&
\delta\bar{C}=0, \\[.15in]
\delta\psi=[\frac{g_{0}}{\nabla^{2}}\partial_{0}\bar{C}]\psi,&
\delta\bar{\psi}=\bar{\psi}\frac{g_{0}}{\nabla^{2}}
\partial_{0}\bar{C}.
\end{array}
\end{equation}
Covariance is not manifest  in the above equations.
The operator $\frac{1}{\nabla^{2}}$ makes
the LM transformation  nonlocal. The LM symmetry leads to
the existence of a nonlocal fermionic Noether current,  and a
corresponding  Noether charge, which generates the LM
 transformation. It also imposes a constraint condition on the physical
states, besides the usual BRST \cite{s6}.

Usually we seek  Poincar\'{e}-covariant symmetries  in gauge theory.
In fact, eq.(2) can be re-expressed in the following
(still not Poincar\'{e}-covariant) form, with the aid of the  equations
of motion for the  $A_{\mu}$ and $C$, namely, on shell,
\begin{equation}
\begin{array}{ll}
\delta A_{\mu}=i\partial_{\mu}(\frac{\partial_{0}}{\nabla^{2}}\bar{C}),
\nonumber\\ [.15in]
\delta C=-\frac{\partial_{0}}{\nabla^{2}}\partial_{\mu}A^{\mu},&
\delta\bar{C}=0, \\ [.15in]
\delta\psi=g_{0}(\frac{\partial_{0}}{\nabla^{2}}\bar{C})\psi,&
\delta\bar{\psi}=\bar{\psi}\frac{g_{0}}{\nabla^{2}}\partial_{0}\bar{C}.\nonumber
\end{array}
\end{equation}
In fact, if choosing the Feynman gauge, i.e., $\xi=1$ in eq.(1),
one can  verify that action is invariant under eq.(3)
 without using the
equations of motion, that is, eq.(3) represents another kind of nonlocal
symmetry existing in QED,  which is equivalent to LM only on shell.
This symmetry is nilpotent and  it  too should  impose a constraint
on  the physical states besides the BRST and LM.
With the interchange  $C\rightarrow i\bar{C}$
and $\bar{C}\rightarrow iC$, one can obtain the anti-form of the
symmetry defined by eq.(3).

LM's work and  eq.(3) show that we do not have the full story  of symmetry
in gauge theory, even in QED.

In this paper we demonstrate that there exists a more general
 Poincar\'{e}-covariant
symmetry  in QED,
which includes  the local and nonlocal symmetries already mentioned.
The symmetry is not nilpotent in general, but it becomes nilpotent under
certain conditions.

In the following we consider  only operators $\hat{\Omega}$
that are sufficiently ``regular" in the sense that they possess  adjoint
$\hat{\Omega}^{\dagger}$ with
\begin{equation}
\int^{+\infty}_{-\infty}d^{3}x\phi(\hat{\Omega}\varphi)=
\int^{+\infty}_{-\infty}d^{3}x(\hat{\Omega}^{\dagger}\phi)\varphi
\end{equation}
under proper boundary conditions of $\phi$ and $\varphi$, in
 which the sign $\dagger$ represents hermitian conjugation.
 Examples: $\partial_{\mu}$, $\nabla^{2}$,
${1}\over{\nabla^{2}}$ \cite{s5}.

The  Lorentz and  Coulomb
 gauges are often used;
their equivalence is easily proved
in path integral form. Being Poincar\'{e} covariant, the Lorentz
gauge is  preferred in path integral formulations,
in view of  eq.(1).
Accordingly, we concentrate
our studies on Poincar\'{e}-covariant symmetries of QED in this paper.
We  consider a Poincar\'{e}-covariant generalization of eq.(2) of the form,

\begin{equation}
\delta A_{\mu}=\partial_{\mu}(fC+g\bar {C}),
\end{equation}
where $f$ and $g$ are  fermionic  operators, that is,
 include  Grassmann constants. In addition,
$f$ and $g$  commute with $\partial_{\mu}$:
\begin{equation}
\partial_{\mu}(f,g)=
(f,g)\partial_{\mu}.
\end{equation}

When $A_{\mu}$ transforms by eq.(5),
the  transformations of
 $C, \bar{C}$
and $\psi$ that  leave the action $S$ invariant are
\begin{equation}
\begin{array}{l}
\delta C=\frac{i}{\xi}g^{\dagger}\partial_{\mu}A^{\mu},\nonumber\\[.15in]
\delta\bar{C}=-\frac{i}{\xi}f^{\dagger}\partial_{\mu}A^{\mu},\\[.15in]
\delta\psi=-ig_{0}(fC+g\bar {C})\psi,\\[.15in]
\delta\bar{\psi}=ig_{0}\bar{\psi}(Cf^{\dagger}+\bar{C}g^{\dagger}),
\end{array}
\end{equation}
in which $f$ and $g$ are regular in the sense of  eq.(4).
One may conclude that even if
eqs.(5,7) define  a nonlocal transformation,  $f$ and $g$ will not  alter the
action
$S$ betweeen  the end points of the integration over space; see for
 example Ref.[4]. Thus  eq.(5) and eq.(7) actually represent a symmetry of QED.

In this generalized  transformation,
the unique requirement on $f$ and $g$ is that they should be regular operators
in
the sense of eq.(4).
 It is easily checked that  the BRST symmetry, the symmetry of eq.(3), and
 their anti-forms are all  special examples of this more general symmetry.

In the following we study some properties of this symmetry.

The generalized symmetry  need not be
 nilpotent in general; see for example  $f=\lambda_{1}, g=\lambda_{2},
 \lambda_{1}\not=\lambda_{2}$.
 Nilpotent symmetries such as BRST define  a  cohomology, but our more
general symmetry does not.
 Moreover, the non-nilpotent transformation defined by eqs.(5,7)
  exhibits the commutation relations of super-Lie algebra.

However, our generalized symmetry is nilpotent under the following conditions.

For  $A_{\mu}$, one can verify that  the following condition leads to
$\delta^{2}A_{\mu}=0$ from eqs.(5,7):
\begin{equation}
fg^{\dag}=gf^{\dag}.
\end{equation}
This condition is evidently  fulfilled in BRST symmetry and  that of eq.(3),
since one of $f,g$  is  zero in those cases.

For $C, \bar{C}$, we see that $\delta^{2}=0$ generally holds only  on shell.
In order to have a ``strong" nilpotency in the theory  in the sense  that
$\delta^{2}(C,\bar{C})=0$  off shell and on,
 we add an auxiliary term $\frac{1}{2}E^{2} $ to the
Lagrangian of eq.(1), where $E$ is a bosonic field. Then,
the transformation
\begin{equation}
\begin{array}{l}
\delta C=\frac{i}{\xi}g^{\dagger}\partial_{\mu}A^{\mu}-\frac{i}
{\sqrt{\xi}}g^{\dagger}E,\\[.15in]
\delta\bar{C}=-\frac{i}{\xi}f^{\dagger}\partial_{\mu}A^{\mu}+\frac{i}
{\sqrt{\xi}}f^{\dagger}E,\\[.15in]
\delta E=\frac{1}{\sqrt{\xi}}\partial_{\mu}
\partial^{\mu}(fC+g\bar{C}),
\end{array}
\end{equation}
fixes the action $S$  (with the auxiliary term  added in)
 and also leads to  $\delta^{2}(C,\bar{C},E)=0$, where
$A_{\mu}$ transforms still according to eq.(5). It is easy to check
that $\delta^{2}A_{\mu}=0$ still holds
under transformation (9).

Thus we have obtained a generalized  symmetry of QED,
represented by eqs.(5,9), which
is relativistically covariant and nilpotent, and  includes
both local and nonlocal forms.

The transformations (5,9) have an evident additive
 group structure. Therefore we take it for granted
that there is an interpolation between the BRST symmetry and that
of eq.(3). Specifically,  if $f$ and $g$ take the values
\begin{equation}
f=\lambda_{1}, g=-i\frac{\partial_{0}}{\nabla^{2}}\lambda_{2},
\end{equation}
one can easily check that eqs.(5,9) express exactly this
interpolation, which is still a nilpotent transformation. We
can construct various symmetries of QED by selecting  $f$ and $g$.

The following
Noether charge  generates
the transformation equations(5,9),
\begin{eqnarray}
Q=i\int d^{3}x\{\partial_{\mu}(fC+g\bar{C})[\partial_{0}A^{\mu}-
(1-\frac{1}{\xi})\partial^{\mu}A_{0}]- \nonumber\\
 \frac{1}{\xi}[g^{\dagger}(\partial_{\mu}A^{\mu}-\sqrt{\xi}E)]
(\partial_{0}\bar{C})-
 \frac{1}{\xi}[f^{\dagger}
(\partial_{\mu}A^{\mu}-\sqrt{\xi}E)](\partial_{0}C)\}.
\end{eqnarray}
If $f,g$ do not depend on  time, then $Q$ becomes
\begin{eqnarray}
Q=\int id^{3}x[-(\partial_{0}\partial_{i}A^{i}-(1-\frac{1}{\xi})\nabla A_{0})
(fC+g\bar{C})- \nonumber\\
(\frac{1}{\xi}\partial_{i}A^{i}-\frac{1}{\sqrt{\xi}}E)
\cdot\partial_{0}(fC+g\bar{C})].
\end{eqnarray}
The nilpotency of the transformations implies $Q^{2}=0$.
The charge is  anti-Hermitian, and  is the foundation of
the cohomology of the generalized  symmetry. Since $f,g$ may be
 operators  generating nonlocal symmetries,  it is useful to extend the usual
cohomology to a nonlocal form. This  work  is not contained in
this paper.

The  physical fields must be invariant
under generalized  symmetry of
 eqs.(5,9). Accordingly, the physical states $|\Psi\rangle$ satisfy
\begin{equation}
\label{e12}
Q|\Psi\rangle=0.
\end{equation}
Evidently, this constraint on the physical states
 covers many special constraints such as BRST and eq.(3)'s.
In this sense, the condition eq.(11) is stronger.

In conclusion, we have exhibited a  relativistically covariant symmetry of  QED
 that covers and generalizes various  local and nonlocal
  symmetries including the eq.(3), BRST and their anti-forms.  This generalized
symmetry  need not be
nilpotent, but becomes nilpotent under a certain condition and
 with the introduction of an auxiliary field. Evidently QED
has new non-nilpotent symmetries. The symmetry imposes
 a constraint on the physical states,   which determines the physical states
 more  strongly than previous  symmetries such as the BRST. We should note
that LM symmetry eq.(2) is not covariant except on shell, so it is not included
in eq.'s(5) and (7) strictly. A larger class of symmetry including covariant
and
non-covariant
forms is worth of investigation.

{\bf Acknowledgement}: The work is supported in part by the Furst Foundation,
 and by the National Science Foundation grant PHY-9114904.
We thank Frank (Tony) Smith, Jr,
for his helpful discussion.

\end{document}